\documentclass[useAMS,usenatbib,letters]{mnras}
\usepackage{amsmath}
\usepackage{amssymb}
\usepackage{graphicx}
\usepackage{natbib}
\usepackage[version=3]{mhchem}
\usepackage{setspace}
\usepackage{float}
\usepackage{mathrsfs}
\title[AGB stars in M4]{An extreme paucity of second population AGB stars in the `normal' globular cluster M4}

\author[B. T. MacLean, S. W. Campbell, G. M. De Silva, et al.]{B. T. MacLean$^{1}$\thanks{E-mail: ben.maclean@monash.edu} S. W. Campbell$^{2,1}$, G. M. De Silva$^{3,4}$, J. Lattanzio$^{1}$, \newauthor V. D'Orazi$^{7,5,6}$, J. D. Simpson$^{3,5,6}$ and Y. Momany$^{7}$ \vspace{2mm}\\
$^{1}$Monash Centre for Astrophysics, School of Physics and Astronomy, Monash University, Victoria 3800, Australia \\ 
$^{2}$Max-Planck-Institut f{\"u}r Astrophysik (MPA), Karl-Schwarzschild-Strasse 1D-85748 Garching, Germany \\ 
$^{3}$Australian Astronomical Observatory, 105 Delhi Rd, North Ryde, NSW 2113, Australia \\
$^{4}$Sydney Institute for Astronomy, School of Physics, The University of Sydney, NSW 2006, Australia \\
$^{5}$Research Centre for Astronomy, Astrophysics \& Astrophotonics (MQAAAstro), Macquarie University, Sydney, NSW 2109, Australia \\ 
$^{6}$Department of Physics and Astronomy, Macquarie University, North Ryde, NSW 2109, Australia \\
$^{7}$INAF Osservatorio Astronomico di Padova, vicolo dell'Osservatorio 5, I-35122, Padova, Italy}

\date{Accepted TBC. Received TBC; in original form TBC}

\begin{document}

\pagerange{\pageref{firstpage}--\pageref{lastpage}} \pubyear{2016}

\maketitle

\label{firstpage}

\begin{abstract}
Galactic Globular clusters (GCs) are now known to harbour multiple
stellar populations, which are chemically distinct in many light
element abundances. It is becoming increasingly clear that asymptotic
giant branch (AGB) stars in GCs show different abundance distributions in
light elements compared to those in the red giant branch (RGB) and
other phases, skewing toward more primordial, field-star-like
abundances, which we refer to as subpopulation one (SP1). As part of a
larger program targeting giants in GCs, we obtained high-resolution
spectra for a sample of 106 RGB and 15 AGB stars in Messier 4 (NGC
6121) using the 2dF+HERMES facility on the Anglo-Australian
Telescope. In this Letter we report an extreme paucity of AGB stars
with [Na/O] $> -0.17$ in M4, which contrasts with the RGB that has
abundances up to [Na/O] $=0.55$. The AGB abundance distribution is
consistent with all AGB stars being from SP1. This result appears to
imply that all subpopulation two stars (SP2; Na-rich, O-poor) avoid
the AGB phase. This is an unexpected result given M4's horizontal
branch morphology -- it does not have an extended blue horizontal
branch. This is the first abundance study to be performed utilising 
the HERMES spectrograph.
\end{abstract}


%\keywords{keywords}
%\maketitle

%%%%%%%%%%%%%%%%%%%%%%%%%%%%%%%%%%%%%%%%%%%%%%%%%%%%%%%%%%%%%%%%%%%%%%%%%%%%%%%

\section{Introduction}
It has been well established that Galactic GCs are typically
homogeneous in the iron peak species \citep{carretta2009}, but are
chemically inhomogeneous in elements affected by proton-capture
reactions (e.g., C, N, O, Na). These inhomogeneities are generally
thought to arise from nucleosynthesis in the first generation of stars
\citep{gratton2004}. Correlations exist in the star-to-star scatter
of some elemental abundances within each cluster, and can be used as tracers of GC
formation (see \citealt{gratton2012} for an extensive review).  One
well-documented chemical pattern is the sodium and oxygen
anti-correlation (Na-O), seen in all GCs \citep{carretta2010}, but not in open 
clusters \citep{desilva2009,maclean2015}. The Na-O
anti-correlation has been documented across both evolved and unevolved
stars in many GCs, indicating that this pattern must be imprinted on the stars at their birth. 
While GC stars can often be separated into more than two
distinct subpopulations in chemical space, for the sake of clarity
here we use just two. Stars with near primordial abundances (Na-poor,
O-rich) we designate as subpopulation one (SP1) and those enriched in sodium and
depleted in oxygen as subpopulation two (SP2). We also define the percentage of RGB and 
AGB stars in a GC that are found to be members of SP2 as $\mathscr{R}_{RGB}$ and 
$\mathscr{R}_{AGB}$, respectively. In studies targeting the RGB in GCs, typical 
$\mathscr{R}_{RGB}$ values are found to be on the order of {$\sim$}60 
\citep[see Figure 16 in][]{carretta2010}.

It is becoming clear that the light element abundance distributions of
AGB stars are significantly different to those of stars in other
phases of evolution in many GCs. \citet{norris19816752} found no
examples of cyanogen (CN) strong AGB stars in NGC 6752 despite the
bimodality of CN strengths in the RGB \citep[see
  also][]{campbell2010}. \citet{campbell2013} observed Na abundances of AGB stars in the same
cluster, and no Na-rich AGB stars were found ($\mathscr{R}_{AGB}=0$, compared 
to $\mathscr{R}_{AGB}=70$). They concluded that
the most likely explanation was that all Na-rich stars (SP2) fail to
reach the AGB phase. M 62 was similarly observed to have a value of $\mathscr{R}_{AGB}=0$ \citep{lapenna2015}, while for 47 Tucanae \citet{johnson2015} found that $\mathscr{R}_{AGB}=37$,
 indicating that a smaller, but significant proportion
of SP2 stars are avoiding the AGB phase. 

We define the `AGB failure rate' $\mathscr{F}$ of a GC to be the percentage of SP2 stars that avoid the AGB (as inferred by its $\mathscr{R}_{\rm RGB}$ value), given by \begin{equation} \mathscr{F} = (1-{\frac{\mathscr{R}_{\rm AGB}}{\mathscr{R}_{\rm RGB}}}){\cdot}100\%, \end{equation}where a value of 100 indicates that no SP2 stars reach the AGB (as in NGC 6752), and a value of zero indicates that $\mathscr{R}_{\rm AGB}=\mathscr{R}_{\rm RGB}$. We provide an
up-to-date summary of this `AGB avoidance' phenomenon in
Table~\ref{tab:agbfailure}.

While theoretical simulations struggle to quantitatively reproduce the
Na distributions of AGB stars in GCs, it likely results from the
He-enrichment of SP2 \citep{charbonnel2013,cassisi2014,charbonnel2015}. 
This results in a smaller envelope mass in the horizontal 
branch (HB) phase, giving rise to higher
surface temperatures. The most extreme of these stars fail to reach
the AGB phase and evolve directly to the white dwarf phase and are known as 
AGB-manqu\'{e} (`failed') stars \citep{agb-manque}. \citet{gratton2010} showed that a large
He-enrichment can result in an extended blue-HB (e.g. NGC 6752
and M 62), suggesting that an extended blue-HB may be indicative of a
high $\mathscr{F}$ value. The recently reported
slight AGB failure rate of 47 Tucanae \citep{johnson2015}, which
contains only a red HB, further supports this link between HB
morphology and AGB avoidance.

The GC Messier 4 (NGC 6121), considered archetypal, is moderately
metal-poor and shows well-populated and distinct red- and blue-HBs
with no significant blue extension
\citep{mochejska2002}. \citet{norris1981m4} first documented the
bimodality of the CN band strength of giant stars in M4 (although we note that \citealp{smithnorris1993} reported a CN-strong monomodality on the RGB), and
\citet{carretta2013} suggested that it only contains two distinct
subpopulations (unlike many GCs which contain three or more). While
the high resolution abundance study of \citet{ivans1999} first hinted
at a disparity between $\mathscr{R}_{RGB}$ and $\mathscr{R}_{AGB}$,
AGB stars have never been systematically studied. M4 has been observed
to show a bimodal distribution in Na and O on the RGB
\citep[hereafter M08]{marino2008} and the HB, with all red-HB stars belonging to SP1 \citep{marino2011}.

In this paper we present results from the first systematic study of the 
AGB of M4, including Na and O abundances for a sample of 106 RGB
stars and 15 AGB stars. This work is part of a larger study
of AGB abundances in GCs \citep{campbell2010,campbell2013}, and presents 
the first abundance results from the HERMES spectrograph on the AAT.

%%%%%%%%%%%%%%%%%%%%%%%%%%%%%%%%%%%%%%%%%%%%%%%%%%%%%%%%%%%%%%%%%%%%%%%%%%%%%%%
%\vspace{-2mm}
\section{Observations and membership}

\begin{figure}
\centering
\includegraphics[scale=0.21]{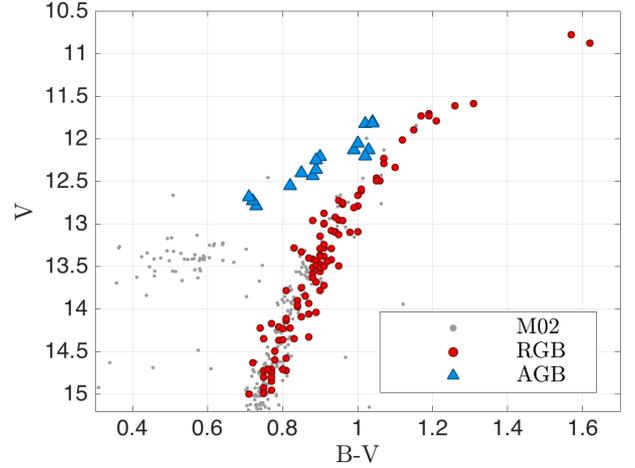}
\caption{Final sample of RGB and AGB stars used in this work are 
  displayed over the larger photometric
  sample of \citet[M02]{mochejska2002}. A value of B$-$V $=+0.05$ 
  was added to the WFI data due to a systematic offset between the two
  photometric data sets.}
\label{fig:cmd}
\end{figure}

For target selection we used photometry of M4 from two sources; UBV
from \citet[with an 8.8'x8.8' field of view]{mochejska2002} and UBVI
from the ESO/MPG Wide Field Imager \citep[WFI, with a 34'x33' field of view;][]{Momany2003}. 
The RGB and AGB were separated
in both V$-$(B$-$V) (see Figure~\ref{fig:cmd}) and U$-$(U$-$I), allowing for
an accurately selected sample of AGB stars. We applied a correction of a constant value $E=0.37$ \citep{hendricks2012} because M4 is affected by
significant reddening.

Spectra were collected in August 2014 and July 2015 using 2dF+HERMES on
the AAT which provides $R=28,000$ resolution spectra in 4 narrow
spectral windows \citep{sheinis2015}. In total 121 targets were
observed with average SNR of 70.  The software package {\sc 2dfdr}
\citep[v6.5]{2dfdr} was used to reduce the spectral data for analysis.

Radial velocities for the HERMES spectra were measured with the
{\sc iraf} \textit{fxcor} package \citep{tody1986}, using a solar
reference template.  We considered all stars with radial velocities above 90 km/s or
below 50 km/s to be non-members. Our average radial velocity after non-member
elimination was {\textless}v{\textgreater} = $70.62 \pm 0.31$ km/s
(${\sigma}=3.45$ km/s), agreeing well with \citet{malavolta2015}, who
report {\textless}v{\textgreater} = $71.08 \pm 0.08$ km/s
(${\sigma}=3.97$ km/s). Individual stellar radial velocities are in
Table~\ref{tab:abunds}. Stellar metallicities (discussed in
$\S$\ref{params}) were used as a further test of cluster membership,
with one AGB star and two RGB stars possessing metallicities that were
farther than 2$\sigma$ from the mean, leaving a sample of 106 RGB and
15 AGB stars. %This number of AGB stars is a substantial fraction of
%${\sim}20$ AGB stars available in the 2 photometric data sets. 
A colour-magnitude diagram of the final sample is presented in
Figure~\ref{fig:cmd}.

%%%%%%%%%%%%%%%%%%%%%%%%%%%%%%%%%%%%%%%%%%%%%%%%%%%%%%%%%%%%%%%%%%%%%%%%%%%%%%%

\section{Method and Results}

\subsection{Atmospheric parameters}\label{params}

BV photometry was used to calculate initial estimates of the
stellar parameters for each star. Effective temperature
($T_{\rm eff}$) was estimated using the calibrated scale of
\citet{ramirez2005}, while surface gravity (log $g$) and
microturbulence ($v_{\rm t}$) were estimated using empirical relations
from \citet{alonso1999} and \citet{gratton1996}, respectively.

Final $T_{\rm eff}$, log $g$ and $v_{\rm t}$ values
(Table~\ref{tab:abunds}) were determined spectroscopically by
measuring the equivalent widths \citep[using the {\sc ares}
  package,][]{ares} of neutral and singly-ionized iron (Fe I \& II,
respectively) absorption lines and calculating the one-dimensional
local thermodynamic equilibrium (LTE) abundance from each line with the
{\sc moog} code \citep[June 2014 release]{moog} and model atmospheres
interpolated from the \citet{atlas9odfnew} grid. Final spectroscopic
parameters were found by requiring excitation and ionisation balance
(with tolerances of 0.015 in slope and 0.1 dex, respectively), as per
\citet{sousa2014} and using our newly developed code {\sc phobos}, to be detailed in MacLean et al. (2016, in preparation). We found the average metallicity of the cluster to
be {\textless}[Fe/H]{\textgreater} = $-1.15 \pm 0.01$
(${\sigma}=0.05$).

%------------------------------------------------------------------------
	
\subsection{Chemical abundances \& Analysis of results}\label{abunds}

We determined LTE abundances for Na and O by measuring the equivalent
widths of a selection of absorption lines. It is well known that many
sodium and oxygen lines deviate from LTE, with systematic offsets that
have been a subject of much research
\citep[e.g.,][]{asplund2005,lapenna2014}. The sodium 568 nm
doublet was measured for each star, and the abundances of each line
were corrected for non-LTE effects as described in \citet{lind2011} by
using the web-based {\sc inspect}
interface\footnote{http://inspect-stars.net}, and adopting the
provided $\Delta$[Na/Fe]$_{\text{\sc nlte}}$ corrections which were
around $-0.15$ dex.

In the case of oxygen, the 777 nm triplet was measured and corrected
for non-LTE effects following \citet{takeda2003oxygen}. Recently,
\citet{amarsi2015} calculated a fine grid of oxygen corrections for
both non-LTE effects and the effects of using 3D
stellar atmosphere models; however the grid range is $T_{\rm eff}$
{\textgreater} 5000K and log $g$ {\textgreater} 3.0; outside the range
of most of our stars.

Final [Na/Fe] and [O/Fe] abundances for all confirmed cluster members
are contained in Table~\ref{tab:abunds}. Also included are uncertainties based on
line-to-line scatter, which are in the range ${\sim}0.10$ to $0.15$ dex. The
abundance sensitivities due to the uncertainty in stellar parameters
are given in Table~\ref{tab:atmos}. These are on the order of 0.02 to
0.15 dex.

%{\begin{table}
%\centering
%\begin{tabular}{cccccc}
%\hline
%Star & Type & {[}O/Fe{]} & {[}O/Fe{]} & {[}O/Fe{]} & {[}O/Fe{]} \\
% & & LTE & A15 & G99 & T03 \\
%\hline
%34336 & RGB & 0.35   & 0.26  & 0.30 & 0.26 \\
%45534 & RGB & 0.72   & 0.60  & 0.66 & 0.59 \\
%47603 & RGB & 0.63   & 0.49  & 0.57 & 0.50 \\
%59950 & RGB & 0.25   & 0.15  & 0.20 & 0.14 \\
%\hline
%\end{tabular}
%\caption{[O/Fe] abundances for four stars with T$_{\text{eff}}\sim$ 5000K and log $g \sim$ 3.0 to highlight the various options for NLTE corrections. Presented are [O/Fe] values calculated using {\sc moog} and assuming LTE, final abundances given by \protect\citet[A15, via the {\sc inspect} interface]{amarsi2015}, and LTE values that have been corrected according to \protect\citet[G99]{gratton1999} and \protect\citet[T03]{takeda2003oxygen}.}
%\label{tab:onlte}
%\end{table}

\begin{table*}
\centering
\caption{Stellar parameters, radial velocities and chemical abundances for each star. Abundance errors reflect line-to-line scatter, and do not take atmospheric sensitivities into account (see Table~\ref{tab:atmos} and text for discussion). Included are the stellar designations used by M08, and the [Na/Fe] and [O/Fe] abundances that they reported. We adopt the \protect\citet{asplund2009} solar abundance values. The full table is available online.}
\label{tab:abunds}
\begin{tabular}{cccccccccccc}
\hline
Star & Type & RV & $T_{\rm eff}$ & log $g$ & $v_{\rm t}$ & [Fe/H] & [O/Fe] & [Na/Fe] & ID \\
& & (km/s) & (K) & (cgs) & (km/s) & & & & M08 \\
\hline
25   & RGB & 66.6 & 5028 & 2.64 & 1.09 & -1.14 $\pm$ 0.12 & 0.30 $\pm$ 0.12 & 0.43 $\pm$ 0.13 & - \\
907  & RGB & 69.4 & 5047 & 2.69 & 0.94 & -1.18 $\pm$ 0.11 & 0.42 $\pm$ 0.12 & 0.37 $\pm$ 0.11 & - \\
1029 & RGB & 72.2 & 4936 & 2.45 & 1.41 & -1.10 $\pm$ 0.09 & 0.09 $\pm$ 0.11 & 0.49 $\pm$ 0.11 & 22089 \\
1129 & RGB & 69.6 & 4886 & 2.20 & 1.20 & -1.11 $\pm$ 0.10 & 0.35 $\pm$ 0.12 & 0.10 $\pm$ 0.10 & - \\
1474 & RGB & 70.4 & 5159 & 2.78 & 0.92 & -1.06 $\pm$ 0.11 & 0.12 $\pm$ 0.12 & 0.39 $\pm$ 0.12 & - \\
\hline
\end{tabular}
\end{table*}

\begin{table*}
\centering
\caption{Abundance uncertainties due to the atmospheric sensitivities of a representative sub-sample of RGB and AGB stars in our M4 data set. Parameter variations (in parentheses) are the expected uncertainties in the respective parameters.}
\label{tab:atmos}
\begin{tabular}{cccccccccccc}
\hline
 &  & & & {[}O/Fe{]} &  &  &  & {[}Na/Fe{]} &  &  &  \\
Star & Type & $T_{\rm eff}$ & log $g$ & $\Delta T_{\rm eff}$ & $\Delta$log $g$ & $\Delta v_{\rm t}$ & Total & $\Delta T_{\rm eff}$ & $\Delta$log $g$ & $\Delta v_{\rm t}$ & Total \\
 & & & & ($\pm$50K) & ($\pm$0.2) & ($\pm$0.1) & & ($\pm$50K) & ($\pm$0.2) & ($\pm$0.1) \\
 \hline
16547 & AGB & 4847 & 1.90 & $\mp$0.07      & $\pm$0.07         & $\mp$0.01    & $\pm$0.10  & $\pm$0.04        & $\mp$0.01        & $\mp$0.02    & $\pm$0.03  \\
16788 & RGB & 3954 & 0.36 & $\mp$0.10      & $\pm$0.11         & $\mp$0.01    & $\pm$0.15  & $\pm$0.05        & $\mp$0.04        & $\mp$0.04    & $\pm$0.02  \\
47603 & RGB & 5251 & 3.01 & $\mp$0.06      & $\pm$0.06         & $\mp$0.01    & $\pm$0.08  & $\pm$0.03        & $\mp$0.02        & $\mp$0.01    & $\pm$0.02  \\
\hline
\end{tabular}
\end{table*}

%%%%%%%%%%%%%%%%%%%%%%%%%%%%%%%%%%%%%%%%%%%%%%%%%%%%%%%%%%%%%%%%%%%%%%

%\subsection{Analysis of results}

\begin{figure}
\centering
\includegraphics[scale=0.17]{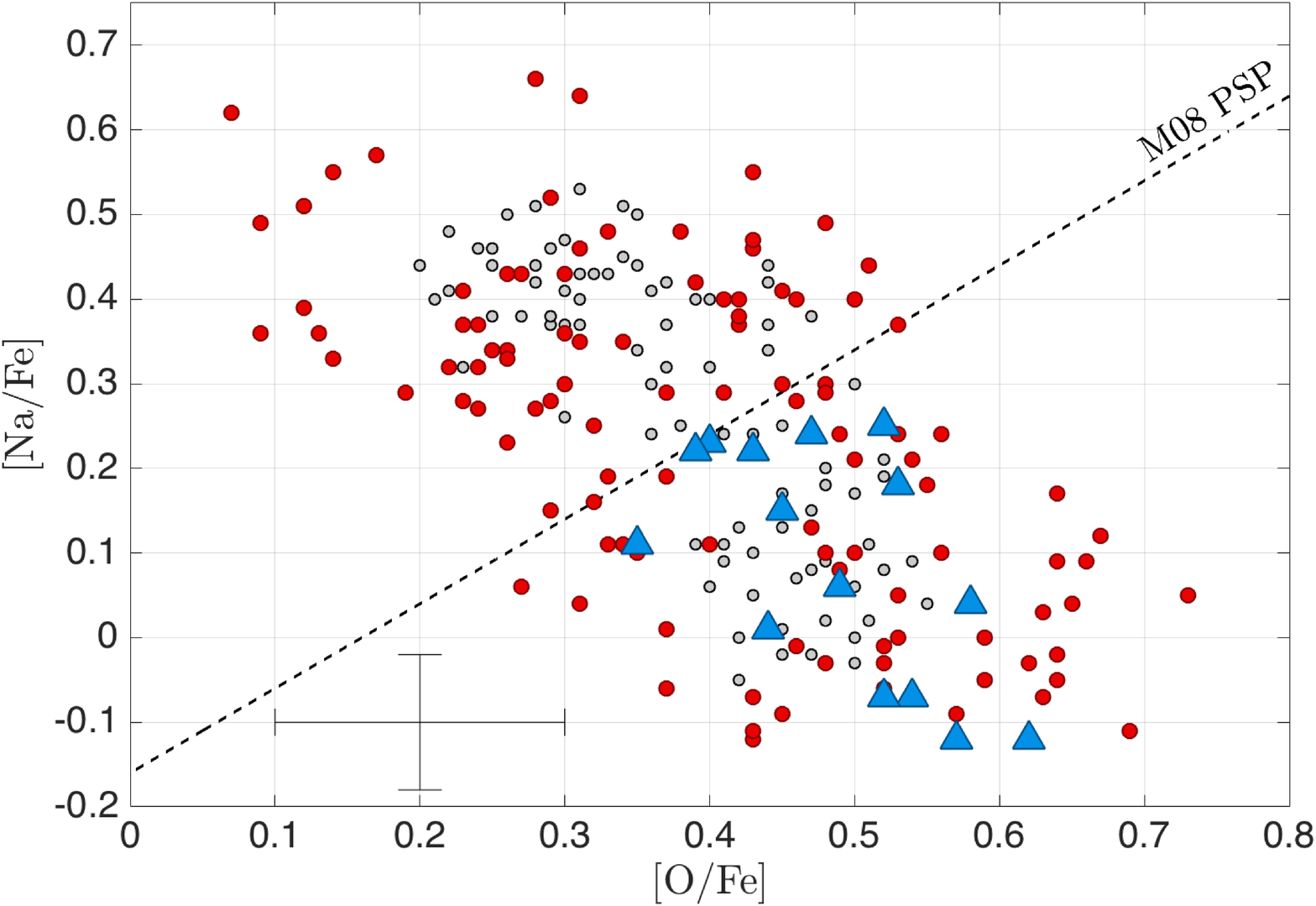}
\caption{Final Na and O abundances for our RGB stars (solid red
  circles) and AGB stars (solid blue triangles; also see CMD in
  Fig.~\ref{fig:cmd}). Shown for comparison is the RGB sample of M08
  (open grey circles). The Na-O anticorrelation is evident. The AGB
  distribution is clearly different from the RGB, showing a paucity of
  SP2 stars. Typical errors in individual abundances are shown in the
  bottom left, while the population separation point (PSP) from M08 is
  indicated by the dashed diagonal line (see text for details).}
%, and its upper- and lower-limits by dashed lines. The shaded
%contours represent the 2D Gaussian kernel density estimation of RGB
%stars, with a bandwidth of 0.10.}
\label{fig:nao}
\end{figure}

There is significant overlap between our RGB sample, that of M08 (51
stars in common), and \citet[hereafter C09, 46 stars in common]{carretta2009vii}. We made
a detailed comparison of this intersecting sample, which revealed that
while there are slight offsets between each study in several stellar
parameters, the scatter among the parameters is consistent with
uncertainties quoted in this work. The results of this comparison are
provided in Table~\ref{tab:overlap}.

%The average differences between the
%parameters of the 51 stars in common between our sample, M08, and C09
%are presented in Table~\ref{tab:overlap}. Both M08 and C09 used the
%[Na/Fe] and [O/Fe] NLTE corrections of \citet{gratton1999}, which in
%both cases were typically smaller than those adopted in this work.

\begin{table}
\centering
\caption{The average differences in parameters and abundances between
  this work, M08, and C09. Uncertainties are the errors on the
  mean. There are no major offsets in the Na and O abundances of our work    
  and that of M08.}
\label{tab:overlap}
\begin{tabular}{ccc}
\hline
Parameter & This study $-$ M08         & This study $-$ C09         \\
\hline
$\Delta T_{\rm eff}$ & 53.9 $\pm$ 1.1        & 154.8 $\pm$ 1.5        \\
$\Delta$log $g$ & $-0.210$ $\pm$ 0.004 & 0.080 $\pm$ 0.003  \\
$\Delta v_{\rm t}$ & $-0.071$ $\pm$ 0.004 & 0.049 $\pm$ 0.007  \\
$\Delta${[}Fe/H{]}  & $-0.092$ $\pm$ 0.001 & 0.054 $\pm$ 0.002  \\
$\Delta${[}O/Fe{]}  & 0.014 $\pm$ 0.003  & 0.197 $\pm$ 0.004  \\
$\Delta${[}Na/Fe{]} & $-0.044$ $\pm$ 0.002 & $-0.126$ $\pm$ 0.003 \\
\hline
\end{tabular}
\end{table}

The [Na/Fe] and [O/Fe] values of our RGB and AGB samples are plotted
along with the RGB sample of M08 in Figure~\ref{fig:nao}. 
The larger scatter in our abundances compared to M08 is due the lower
signal-to-noise ratio of our data. We attempted to define a population
separation point (PSP) in our RGB sample by identifying a minimum
between the two subpopulations (see Fig. 7 in M08; however the
uncertainties in our abundances combined with the relatively small
spread in Na and O in M4 did not allow us to define one reliably. We
have instead included the M08 PSP at [Na/O] $=-0.16$ in the
figure. Using this PSP we find $\mathscr{R}_{\rm RGB}=55$, which is consistent with that found by M08. It is also
close to $\mathscr{R}_{\rm RGB}=62{\pm}4$, as determined for the double main sequence using photometric star counts \citep{milone2014}.

The usual Na-O anticorrelation can be seen in the RGB sample, with a
spread of ${\sim}0.8$ dex in [Na/Fe] and ${\sim}0.6$ dex in [O/Fe]. In
contrast, the AGB distribution is heavily skewed to SP1 compositions,
with the spread in AGB abundances being restricted to ${\sim}0.4$ dex
in [Na/Fe] and ${\sim}0.3$ dex in [O/Fe]. There are no AGB stars 
above the M08 PSP, giving $\mathscr{R}_{\rm AGB}=0$ and $\mathscr{F}=100$.

%%%%%%%%%%%%%%%%%%%%%%%%%%%%%%%%%%%%%%%%%%%%%%%%%%%%%%%%%%%%%%%%%%%%%%%%%%%%%%%
%\clearpage
%\vspace{-2mm}
\section{Discussion and conclusions}

\begin{table*}
\centering
\caption{A summary (including our current M4 results) of the subpopulation membership percentages of RGB and AGB stars in GCs as reported in the literature. Also included are metallicities, HB-morphology (`R' indicates the presence of a red-HB, `B' a blue-HB, and `EB' an extended blue-HB), and the elemental inhomogeneities used to separate the subpopulations. Sample sizes are the total number of RGB and AGB stars analysed in each respective study.}
\label{tab:agbfailure}
\begin{tabular}{ccccccccccc}
\hline
NGC  & Other  & {[}Fe/H{]} & $\Delta$Y & \multicolumn{2}{c}{Sample size} & $\mathscr{R}_{\rm RGB}$ & $\mathscr{R}_{\rm AGB}$ & $\mathscr{F}$ & HB & Elements     \\
 & & & & RGB & AGB & & & & morphology & used  \\
\hline
104  & 47 Tuc & $-0.68$  &  0.02\footnotemark & 113\footnotemark & 35 & 55 & 37 & 33 & R        & Na           \\
5272 & M 3    & $-1.50$  &  0.02\footnotemark & 46\footnotemark   & 9  & 50 & 33 & 34 & R+B      & Al            \\
5904 & M 5    & $-1.29$  & -  & 107$^5$  & 15 & 50 & 33 & 34 & R+B+EB   & Al            \\
6121 & M 4    & $-1.15$  &  0.01\footnotemark & 106\footnotemark  & 15 & 55 & 0 & 100 & R+B      & Na-O         \\
6205 & M 13   & $-1.53$  &  0.06\footnotemark & 67$^5$   & 14 & 70 & 27 & 61 & B+EB     & Al           \\
6266 & M 62   & $-1.10$  &  0.08\footnotemark & 13\footnotemark   & 5 & 46 & 0 & 100 & R+B+EB   & Na-O, Mg-Al   \\
6752 & -      & $-1.54$  &  0.03\footnotemark & 24\footnotemark   & 20 & 70 & 0 & 100 & EB     & Na           \\
7089 & M 2    & $-1.65$  &  0.07\footnotemark & 12$^5$   & 5 & 80 & 40 & 50 & B+EB     & Al           \\
\hline
\multicolumn{11}{l}{\footnotesize{$^2$ \protect\citet{milone201247tuc}} $^3$ \protect\citet{johnson2015} $^4$ \protect\citet{valcarce16} $^5$ \protect\citet{garciahernandez2015}} \\
\multicolumn{11}{l}{$^6$ \protect\citet{valcarce2014} $^7$ This study $^8$ \protect\citet{dalessandro2013} $^9$ \protect\citet{milone2015} $^{10}$ \protect\citet{lapenna2015}} \\ 
\multicolumn{11}{l}{$^{11}$ \protect\citet{milone2013} $^{12}$ \protect\citet{campbell2013} $^{13}$ \protect\citet{milone2015m2}}
\end{tabular}
\end{table*}

%We have reported radial velocities, atmospheric parameters, and
%abundances for Fe, Na and O for a sample of 106 RGB stars and 15 AGB
%stars in M4. The RGB stars show a Na-O anti-correlation and an SP1:SP2
%ratio of approximately XX:YY, both typical of Galactic GCs and
%consistent with previous RGB studies of this cluster.

%and an average [Na/O] abundance of $-0.16 \pm 0.03$ (${\sigma}=0.31$). 

The novel feature of this work is the AGB sample. This is the first
time that the AGB has been specifically targeted in M4. We found that
our sample of AGB stars has a much smaller spread in Na and O abundances
than the RGB sample. Following the population separation point from M08, the
AGB distribution is consistent with $\mathscr{R}_{\rm AGB}=0$ and $\mathscr{F}=100$. 
However, given (i) that the tails of the SP1
and SP2 RGB distributions appear to overlap in [Na/O] (cf. Fig. 7 in
M08), and (ii) the uncertainties in our data, it is possible that the
higher-Na (lower-O) AGB stars actually lie in the tail of the SP2
distribution. This would increase the value of $\mathscr{R}_{\rm AGB}$ from zero. Thus, until better data are obtained for the AGB stars, some uncertainty
remains as to the exact failure rate ($\mathscr{F}$) of M4. It is clear however that the
majority of AGB stars in M4 have compositions typical of SP1 stars
(Fig.~\ref{fig:nao}). A further uncertainty lies in the NLTE
corrections, which may not be accurate for AGB stars. This was
suggested by \citet{lapenna2015} as a possible risk to determining
subpopulation membership based on NLTE-affected Na lines; however
there is growing evidence from a number of studies
(including the \citealp{lapenna2015} M 62 study) that SP2 AGB avoidance
is common in GCs. These studies are based on various elements and
atomic lines (Table~\ref{tab:agbfailure}).

% they did confirm a lack of SP2
%AGB stars in M 62 by using both Na and O (using the forbidden
%[\ion{O}{i}] line at 6300.3 \AA, which is unaffected by NLTE
%effects). There is also a growing number of independent studies that
%report reduced SP2 stars on the AGB, based on various elements and
%atomic lines (Table~\ref{tab:agbfailure}).

To put this finding in context we provide a summary of AGB
and RGB subpopulation membership in
Table~\ref{tab:agbfailure} for the GCs for which $\mathscr{F}$ values have been determined. 
The table also includes HB morphology
descriptions. As previously mentioned, recent observational and
theoretical work has suggested a close link between HB morphology, 
He-enrichment, and $\mathscr{R}_{\rm AGB}$ values in
GCs. For example, helium enrichment in NGC 6752 and M 62 -- both of
which have $\mathscr{F}=100$ -- has been inferred to
be relatively high, with ${\Delta}$Y $\simeq$ 0.03 and 0.08,
respectively \citep{milone2013,milone2015}. Both these GCs also have
extended blue HBs. In Table~\ref{tab:agbfailure} the GC with the
closest HB morphology to M4 is M3. The helium spread in M3 has been
reported to be up to ${\Delta}$Y $\sim 0.02$ \citep{valcarce16}. In
terms of AGB stars, \citet{garciahernandez2015} report that M3 has $\mathscr{F}=34$, as is
(qualitatively) expected from its HB morphology and moderate He
enrichment. Given M4's low He enrichment \citep[${\Delta}$Y $ \simeq 0.01$,][]{valcarce2014}, 
and its lack of an extended blue-HB, it would be
expected that the AGB abundance distribution should be similar
to M3 or 47 Tucanae (red-HB only, $\mathscr{F}=33$). It should be 
noted that age is a critical parameter in HB morphology, and that the differences in ages 
between these three clusters (M3, M4 and 47 Tuc) are up to ${\sim}1.2$ Gyr \citep{carretta2010,charbonnel2015}. 
Instead of showing a low to moderate AGB failure rate, as may be expected, 
M4 is consistent with a GC with an extended blue-HB and a
higher SP2 He abundance. Furthermore, a comparison between the HB
morphologies of M4 and NGC 6752 shows that the M4 blue HB ends
approximately where the NGC 6752 HB starts (around Teff ${\sim}7000$
K). Using star counts \citet{campbell2013} report that it is only the
stars hotter than ${\sim}11500$ K (the Grundahl Jump) that fail to
reach the AGB, ie. far beyond the bluest HB stars in M4. Models
predict AGB avoidance only at even higher temperatures (see eg. Fig. 3
in \citealt{campbell2013}). This suggests that there is one (or more)
extra parameters that determine AGB avoidance, and that the HB stellar
models cannot reproduce the observations, particularly for M4.

The extreme paucity (or possible total lack, $\mathscr{F}=100$) of SP2 AGB stars in the
`normal' globular cluster M4 imposes further constraints upon the
theory of the evolution of low-mass metal-poor stars, in particular
the evolution of SP2 stars through the giant phases of evolution, and
how this may be tied to their initial He abundances, mass-loss
histories, and other factors. Finally, this result (i) demonstrates that star counts using the AGB to test stellar evolution time-scales may be unreliable because of altered CMD 
number statistics, (ii) could help to understand the source of excess 
UV flux in the spectra of elliptical galaxies due to the high surface 
temperatures of AGB-manqu\'{e} stars, and (iii) may provide
indirect clues to the formation history of globular clusters and their
HB morphologies.

%\clearpage
%%%%%%%%%%%%%%%%%%%%%%%%%%%%%%%%%%%%%%%%%%%%%%%%%%%%%%%%%%%%%%%%%%%%%%%%%%%%%%%
%\vspace{-3mm}
\section*{Acknowledgements}

Based in part on data acquired through the AAO, via programs 14B/27 and 15A/21 (PI Campbell). VD acknowledges support from AAO distinguished visitor program 2016.

%----------------------------------------------------------------------------------------------------------------------
\vspace{-5mm}
\bibliography{References}

%----------------------------------------------------------------------------------------------------------------------

\label{lastpage}

\end{document}